\documentclass[11pt]{amsart}
\usepackage{geometry}            
\usepackage[parfill]{parskip} 
\usepackage{graphicx}
\usepackage{amssymb}
\usepackage{epstopdf}
\title{Climate Change and Open Science}
\author{Ian Percival}                                                                                                                                                   \date{2013aug23}
\begin{document}
\maketitle        

\bibliography{0}

\begin{thebibliography}{99}

\bibitem{IPCC} IPCC, \emph{Climate Change 2007: Synthesis Report. Contribution of Working Groups I, II and III to the Fourth Assessment Report of the Intergovernmental Panel on Climate Change} [Core Writing Team, Pachauri, R.K and Reisinger, A. 
(eds.)]. IPCC, Geneva, Switzerland, 104 pp.
\bibitem{stern} Nick Stern, \emph{Stern Review on the Economics of Climate Change 2008: Executive Summary, H.M.Treasury} (nationalarchives.gov.uk).
\bibitem{liebenberg} Louis Liebenberg, {\em The Art of Tracking, The Origin of Science}, David Philip Publishers (Pty) Ltd. 208 Werdmuller Centre, Claremont 7700, South Africa. (1990).
\bibitem{moore} Gordon Moore, {\em Progress in Digital Integrated Electronics} IEEE, IEDM Tech Digest  p11-13 (1975).
\bibitem{ostrom} Charlotte Hess and Elinor Ostrom (eds) {\em Understanding Knowledge as a Commons} MIT Press (2011).
\bibitem{suber} Peter Suber in [5].
\bibitem{schweik} Charles M. Schweik  in [5].
\bibitem{suber1} Peter Suber, {\em Open Access} MIT Press (2012).

\end{thebibliography}
\bibliographystyle{alpha}

\emph{\bf Sections}

1. Climate change 

2. Origin of science 

3. The paper era

4. The silicon era: internet science

5. Conclusion

.

{\bf 1. Climate change} 

The world economy is now big enough to make significant and rapid changes in our environment and of life on this planet ~\cite{IPCC}. So whether we like it or not, we humans have become custodians of the planet: to look after it well, we need to take appropriate action, which depends on finding reliable answers to the following questions:
 
\emph{Q1 Given the current trajectory of atmospheric carbon-loading, and the resultant climate change, how can we sustainably protect the biosphere?}

\emph{Q2 Given this climate change, how can we make sure that our descendants will have better lives than ours?}

The answers need to be based on sufficiently reliable physical, biological and social science, leading to remedial action within the next few decades ~\cite{stern}, with negligible probability of failure, so we need to accelerate progress in the relevant sciences by many orders of magnitude. At first sight this appears to be a hopeless task, but closer investigation gives hope. 

.

{\bf 2. Origin of Science.} 

Charles Darwin and Louis Liebenberg have a lot in common. Their early research was supported financially by their parents, and both studied origins: Darwin's book on the origin of species was first published in 1859; Liebenberg's book ~\cite{liebenberg} on the origin of science was published in 1990 and is now freely available on the internet. Both risked their lives for their work. 

Discoveries were made in prehistory before they were recorded.  Useful discoveries were retained and developed in small communities. Their application contributed to the fitness of those communities and helped them to survive, so it was subject to natural selection: the theory of natural selection applies to the prehistory of discovery.  For example, the discoveries of the animal trackers of the Kalahari desert in South Africa were described by Liebenberg in his book from first-hand experience. According to him, tracking involved creative problem-solving in which hypotheses were tested against evidence, rejecting those which do not stand up and replacing them with better ones, a significant but originally unrecorded contribution to scientific method. Unlike Liebenberg in his book, I think that the transition from localized discoveries in small communities to what we now call science took place only after records were invented.   The invention of records was an advance in technology, whose impact was not restricted to science. With scientific records, artificial selection of scientific discoveries supplements natural selection, accelerating scientific progress. The records are transmitted widely between communities in both space and time, allowing thorough testing by observation and experiment. Once scientific records were invented, the interaction between technology and science became a permanent feature of both. Today the application of science makes an important contribution to the fitness and survival of human communities and to the future of humanity as a whole.

.

{\bf 3. The paper era}

Darwin's scientific records were written by hand on paper and published in print on paper. In his time, rapid communication was by post or private courier, lasting records were kept in published books and journal articles: this was during the \emph{paper era} of science, during which the tradition began of referring to past publications, producing a time-dependent network of references, each branch of the network linking two scientific documents or `papers', the later citing the earlier. The network was very sparse near the beginning of the era, because there were few earlier paper documents which could be referenced. By contrast, near the end of the paper era in the 20th century, the network grew so much that fields of science became increasingly tribal. Each field and sub-field of science developed its own exclusive language, slowing down the transmission of scientific records between them and consequently slowing scientific progress too. 

.

{\bf 4. The silicon era: internet science}

Paper is now being replaced: most new scientific records are silicon-based. The internet has already broken down many of the barriers between the sciences; the process is not complete, but is likely to continue until all the barriers are down, making science a seamless whole. Individual scientists and small local communities of scientists are still the basic units of scientific progress, but these days finding answers to some of the most important questions, like Q1 and Q2 above, needs communication and collaboration across distant parts of the globe \emph{and} between traditionally distant fields of science. For this communication the barriers \emph{must} come down. In order to satisfy the time constraint of Q1 and Q2, remedial action within a few decades, progress in relevant  internet science needs to grow exponentially, with an exponent as large as possible, by analogy with, and dependent on, the exponential growth of electronics according to Moore's Law ~\cite{moore}. This can be optimized only by bringing in as many internet scientists as possible. For most scientists, access to the internet is not enough to participate in world science, because at present doing science on the internet needs money. This problem can be overcome by making the relevant science open-access (OA). Open access  is defined by Peter Suber in an article in the book ``Understanding Knowledge as a Commons'' ~\cite{ostrom} as ``not only free of charge to everyone with an internet connection, but free of most copyright and licensing restrictions'' ~\cite{suber}. In the same book, an article by Charles Schweik ~\cite{schweik} describes how OA ``has the potential to lead to more rapid progress in research than is possible within the existing structure of scientific research and publication''. Suber has also written an excellent book on OA, which is itself open access \cite{suber1}. 

.
 
{\bf 5 Conclusion}

If we are to answer the questions Q1 and Q2 of the first section in time, and overcome the problems of climate change, we need a new kind of science. The keys to this science are open access and cooperation, which gives an added urgency to the open access program.   

.

{\bf Acknowledgements}  {\emph{Thanks} to Tim Palmer, Michelle Pavitt, Dave Pick, David Waxman, David Weir, Paul Williams.}

I am at the Physics and Astronomy Departments of the University of Sussex and Queen Mary, University of London.

\end{document}